# Stable glasses of organic semiconductor resist crystallization

*Kushal Bagchi* [a], *Marie E. Fiori* [a], *Camille Bishop* [a], *M.F. Toney*[b,c] *and M.D. Ediger*[a*]

[a] Department of Chemistry, University of Wisconsin-Madison, Madison, Wisconsin 53706, United States

[b]Stanford Synchrotron Radiation Lightsource, SLAC National Accelerator Laboratory, Menlo Park, California 94025, United States

[c]Department of Chemical and Biological Engineering, University of Colorado Boulder, Boulder, CO 80309, United States

AUTHOR INFORMATION

Corresponding Author: M.D. Ediger

*Correspondence to: M.D. Ediger

Email : ediger@chem.wisc.edu

**ABSTRACT:**

The instability of glassy solids poses a key limitation to their use in several technological applications. Well-packed organic glasses, prepared by physical vapor deposition (PVD), have drawn attention recently because they can exhibit significantly higher thermal and chemical stability than glasses prepared from more traditional routes. We show here that PVD glasses can also show enhanced resistance to crystallization. By controlling the deposition temperature, resistance towards crystallization can be enhanced by at least a factor of ten in PVD glasses of the model organic semiconductor Alq3 (Tris(8-hydroxyquinolinato) aluminum). PVD glasses of Alq3 first transform into a supercooled liquid before crystallizing. By controlling the deposition temperature, we increase the glass→liquid transformation time thereby also increasing the overall time for crystallization. We thus demonstrate a new strategy to stabilize glasses of organic semiconductors against crystallization, which is a common failure mechanism in OLED (organic light emitting diode) devices.

## INTRODUCTION:

Glasses are disordered and non-equilibrium solids that are traditionally formed by cooling a supercooled liquid[1,2]. Apart from being interesting from a fundamental physics perspective[3], glassy solids are also used in a broad range of technological applications, from electrical transformers, that use metallic glasses[4] to OLED (organic light emitting diode) devices that use molecular glasses[5,6,7]. While glasses offer several advantages over crystalline solids, such as macroscopic homogeneity (lack of grain boundaries) and compositional flexibility (dopants can be easily dispersed inside a glassy solid), a key limitation of glasses is their long-term stability[8]. Glasses physically age over time[9,10] and can crystallize[11,12]. Additionally, glasses can be more prone to chemical degradation than crystals[8]. The physical and chemical stability of molecular glasses has been shown to be vital to the device lifetime of OLEDs[6,13].

Physical vapor deposition (PVD) can be used to form highly stable glasses[14,15,16]. This is significant technologically, as PVD is the standard route to prepare thin glassy films of organic semiconductors for OLED (organic light emitting diode) applications. The kinetic stability, as well as other properties of PVD glasses, are highly sensitive to the deposition temperature and rate[17,18,19]. The most stable PVD glasses are formed at a slow deposition rate and a deposition temperature that is 75-95% of the glass transition temperature. When annealed above the glass transition temperature, the most stable glasses formed by PVD can take up to ~ $10^5$ times longer to transform into a supercooled liquid than an ordinary liquid-cooled glass[20]. This enhanced kinetic stability in PVD glasses is often also accompanied by superior chemical stability. Relative to liquid-cooled glasses, stable PVD glasses have been shown to be more resistant to light-induced reactions[21,22], reactions with atmospheric gases[23], and water vapor uptake[24]. Depositing organic semiconductor layers (the emissive and electron transport layers), at temperatures that produce the most stable glasses was shown to increase the device lifetime of an OLED by a factor of five, relative to deposition at room temperature[6] (which resulted in a less stable glass).

While well-established strategies exist to increase the kinetic and chemical stability of PVD glasses, the same is not true for stability against crystallization. Preparing glasses that are stable with respect to crystallization is important for a broad range of applications. For instance, a common failure mechanisms for OLED devices is crystallization of the glassy charge transport and emissive layers[25,26]. It has been shown that crystallization of the common electron transport layer Alq3 (Tris(8-hydroxyquinolinato) aluminum) in OLEDs leads to cathode delamination which in turn causes the formation of non-emissive dark spots[26]. For vapor-deposited organic glasses, there is only one example in the literature (to the best of our knowledge) where glass stability has been shown to influence crystallization kinetics; crystallization was shown to be ~ 30% slower in a stable PVD glass of celecoxib than for its liquid-cooled glass[27].To test the generality and applicability of this result, crystallization studies need to be performed on a larger variety of glass formers, including organic semiconductors, and also across a broader range of conditions. Using glass stability to prevent or delay crystallization could have importance implications for organic electronic devices such as OLEDs.

In this manuscript, we investigate the role of deposition temperature on the crystallization of vapor-deposited glasses of Alq3, a molecule which has been used in OLED devices for several decades[28], and is also of interest for organic lasers[29] and spintronics[30]. We show that Alq3 glasses prepared at different deposition temperatures can vary by at least a factor of ten in crystallization time. We show that crystallization of PVD glasses of Alq3 (upon annealing) follows a two-step process whereby the glass first transforms into a supercooled liquid and crystallizes thereafter. PVD glasses of Alq3 deposited at the optimal deposition temperature take longer to transform into a supercooled liquid and therefore also crystallize more slowly. Our study shows that like chemical and kinetic stability, the stability against crystallization in PVD glasses can also be modulated by at least an order of magnitude by controlling the substrate temperature during deposition.

**METHODS:**

*Sample Preparation***:** Alq3 (99.995% purity, trace metals basis) was purchased from Sigma Aldrich and used without further purification. The films were deposited on <100> cut silicon wafer that had ~2 nm of native oxide. PVD (physical vapor deposition) was performed in a vacuum chamber with base pressure of $\sim 10^{-6}$ Torr. The films were deposited at a rate of 0.15-0.2 nm/s. Deposition rate was monitored during deposition with a quartz crystal microbalance. The films had a thickness of 450-600 nm. Film thickness was measured after deposition with variable angle spectroscopic ellipsometry (VASE).

*GIWAXS:* GIWAXS measurements were performed in Beamline 11-3 at the Stanford Synchrotron Radiation Lightsource (SSRL). The X-ray wavelength for the measurements was 0.973 Å. The sample to detector distance was set at 315 mm. The exposure time for each measurement was 10 s. Data were collected at an incidence angle of 0.14°, which is above the critical angle of Alq3, and therefore representative of the bulk structure of the thin film. A "$\chi$ correction"[31] was applied to the raw diffraction patterns to produce the images in **Figure 1** and for the subsequent analysis in **Figure 2-4.** For all our analysis, we define $\chi$ to be 0° along $Q_z$ and 90° along $Q_{xy}$. To produce the plots in **Figure 2 and 3,** intensity from $\chi$ of 10° to 85° was summed; a linear background subtraction was performed to the raw I(Q) vs Q profiles subsequently.

The degree of crystallinity is determined by calculating the area under the crystalline peaks relative to the total scattered area. The degree of crystallinity is expressed mathematically as:

$$Degree\ of\ crystallinity = \frac{Total\ area - amorphous\ area}{Total\ area} = \frac{Area\ under\ crystalline\ peaks}{Total\ area} \quad (1)$$

The calculation of degree of crystallinity is outlined in detail in the supplementary information (see **Figure S6** and **S7**). To quantify the uncertainty in the degree of crystallinity, as calculated above, we compare it with an alternate method in section VI of the supplemental information. The agreement between the two different methods is good.

To quantify scattering anisotropy (used below to monitor the transformation of the as-deposited glass into the supercooled liquid), we utilize the Hermans order parameter, $S_{GIWAXS}$[32]. The angular distribution of scattered intensity is used to compute the order parameter using the equations:

$$S_{GIWAXS} = \frac{1}{2}(3 < \cos^2\chi > -1) \quad (2)$$

where

$$< \cos^2\chi > = \frac{\int_0^{90} I(\chi)(\cos^2\chi)(\sin\chi)d\chi}{\int_0^{90} I(\chi)(\sin\chi)d\chi} \quad (3)$$

Equation 2 has the mathematical form typical of order parameters based on the second order Legendre polynomial. Equation 3 incorporates the sine correction which is required for quantitative assessment of scattering anisotropy in 2D GIWAXS measurements[33]. The following steps were performed to calculate the order parameter 1) For a sample produced at each deposition temperature the peak position for the layering feature near 0.8$Å^{-1}$ was evaluated 2) Data was summed from Q of 0.75 to 0.85 $Å^{-1}$, 0.73 to 0.83 $Å^{-1}$ and 0.71 to 0.81 $Å^{-1}$ for the glasses deposited at 240 K, 280 K and 340 K, respectively. The integration slice was chosen such that the peak center is at the middle of the slice. 3) The I(χ) vs χ plot, obtained in step 2, was fit to a polynomial, and extrapolated to fill in the data in the missing angles. 4) Data from Q of 2.17 to 2.27 $Å^{-1}$ was used to evaluate the background, and subtract the curve obtained in step 3. 5) The background subtracted $I(\chi)$ vs $\chi$ curve was used to evaluate $S_{GIWAXS}$ (Hermans order parameter) using equations 2 and 3.

**RESULTS AND DISCUSSION:**

To understand the influence of deposition temperature on crystallization kinetics, we perform in-situ GIWAXS (grazing incidence wide angle x-ray scattering) on Alq3 glasses prepared at different deposition temperatures. We study films prepared at 240 K ($0.54T_g$) and 340 K ($0.76T_g$), as these deposition temperatures are expected to yield glasses with low and high kinetic stability, respectively[34]. The films are annealed at 453 K, which is ≈ 5 K above the reported glass transition temperature of Alq3[35]. Shown in **Figure 1** are GIWAXS patterns collected at various annealing times for Alq3 glasses deposited at 240 K (a-d) and 340 K (e-h). In a GIWAXS pattern, $Q_z$ and $Q_{xy}$ are the out of plane and in-plane scattering vectors, respectively[36]. The colors represent intensity, as shown by the scale bar in **Figure 1**. Both the as-deposited glasses ($T_{sub}$=240 K and 340 K, t=0 s) in **Figure 1** exhibit broad scattering features at ≈ 0.8 $Å^{-1}$ and ≈ 1.6 $Å^{-1}$; such broad features are characteristic of amorphous materials. After 210 s of annealing, the diffraction pattern from the glass deposited at 240 K exhibits several sharp rings, which is characteristic of polycrystalline materials. On the other hand, the Alq3 glass deposited at 340 K is largely amorphous even after 2610 s of annealing. **Figure 1** provides direct qualitative evidence that an Alq3 glass deposited at 240 K crystallizes *at least ten times* faster than a film deposited at 340 K. (A representative time-temperature profile for these measurements is shown in **Figure S2**.)

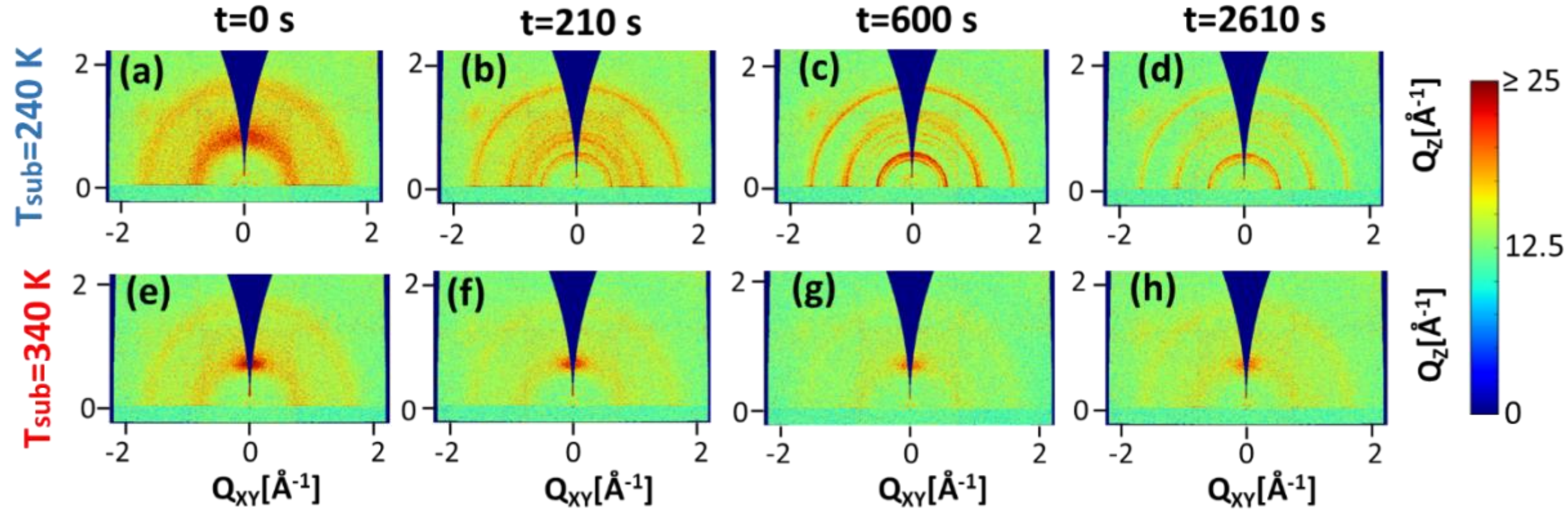

**Figure 1:** Two-dimensional X-ray scattering patterns for Alq3 glasses deposited at 240 K(a-d) and 340 K(e-h) as a function of annealing time at 453 K, which is ≈5 K above the glass transition temperature of Alq3. While the glass deposited at 240 K shows sharp crystalline features by 210 s, the glass deposited at 340 K is largely amorphous after 2610 s of annealing.

To assess the onset and progress of crystallization in PVD glasses of Alq3 we evaluate the 1D scattering plots shown in **Figure 2**. These 1D plots are obtained by summing intensity over the azimuthal angle in a 2D GIWAXS pattern. Shown in **Figure 2** are 1D plots for Alq3 films deposited at 240 K and 340 K, at various annealing times. The t=0 s pattern in **Figure 2** shows that the as-deposited glasses of both the $T_{sub}$=240 K and $T_{sub}$=340 K films are fully amorphous; thus, PVD glasses of Alq3 only crystallize upon annealing after deposition. **Figure 2a** shows that there is a lag period of 90 s for the $T_{sub}$=240 K sample, after which it crystallizes. After 330 s of annealing, crystallization is nearly complete in this sample; there is no significant evolution in the diffraction pattern after this time. The glass prepared at 340 K crystallizes much more slowly. Small sharp crystalline features between 0.4 and 0.65 $Å^{-1}$ are visible in the t=2100 s pattern, and not in the t=1770 s pattern. The onset time for crystallization in the $T_{sub}$=340 K sample (roughly 2000 s) is an order of magnitude larger than the crystallization onset time for the glass deposited at 240 K. After a prolonged annealing time, even the $T_{sub}$=340 K sample crystallizes significantly, as evident from the t=2940 s pattern in **Figure 2B.** It can be seen in **Figure 2** that both the $T_{sub}$=240 K and $T_{sub}$=340 K eventually form the same polymorph upon annealing. Consistent with previous studies[37], we find that the alpha polymorph forms upon annealing above $T_g$ (see **Figure S5**).

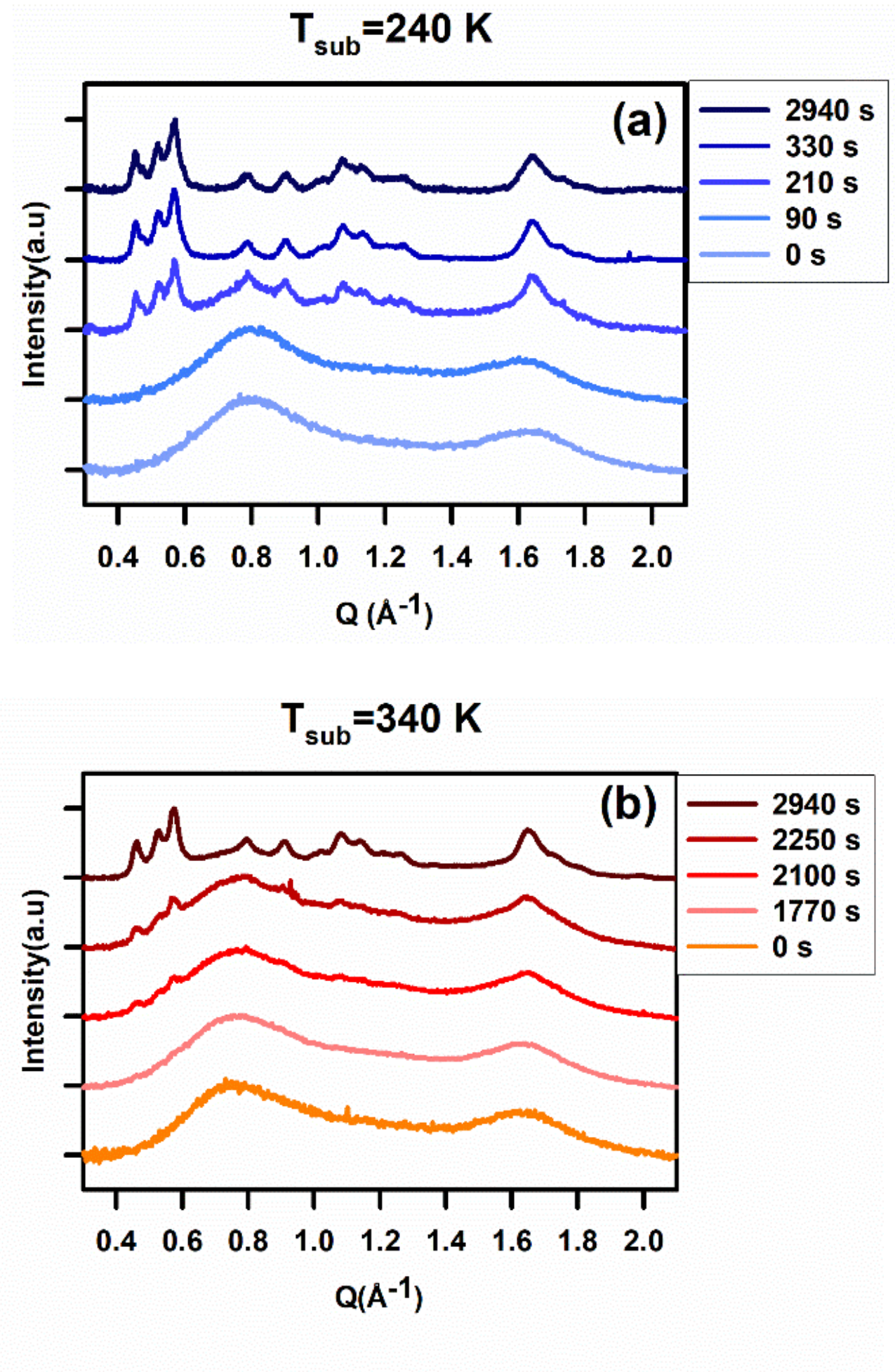

**Figure 2**: X-ray scattering profiles for Alq3 glasses deposited at 240 K (a) and 340 K (b) at various annealing times. The Alq3 glass deposited at 240 K and annealed (at 453K) for 210 s is clearly far more crystalline than the glass deposited at 340 K and annealed for 2100 s (at 453K). Thus, the glass deposited at 340 K is at least ten times more resistant towards crystallization than the glass deposited at 240 K.

We now turn our attention to quantitatively evaluating crystallization kinetics. Shown in **Figure 3** is the degree of crystallinity as a function of annealing time. **Figure 3** shows that while the Alq3 glass deposited at 240 K finishes crystallizing by ~420 s, the $T_{sub}$=340 K glass shows no evidence of crystallinity even after 1800 s of annealing. The glass deposited at 280 K crystallizes at an intermediate rate. The degree of crystallinity is mathematically defined in the methods section; a detailed discussed of this procedure can be found in section V of the supplemental information.

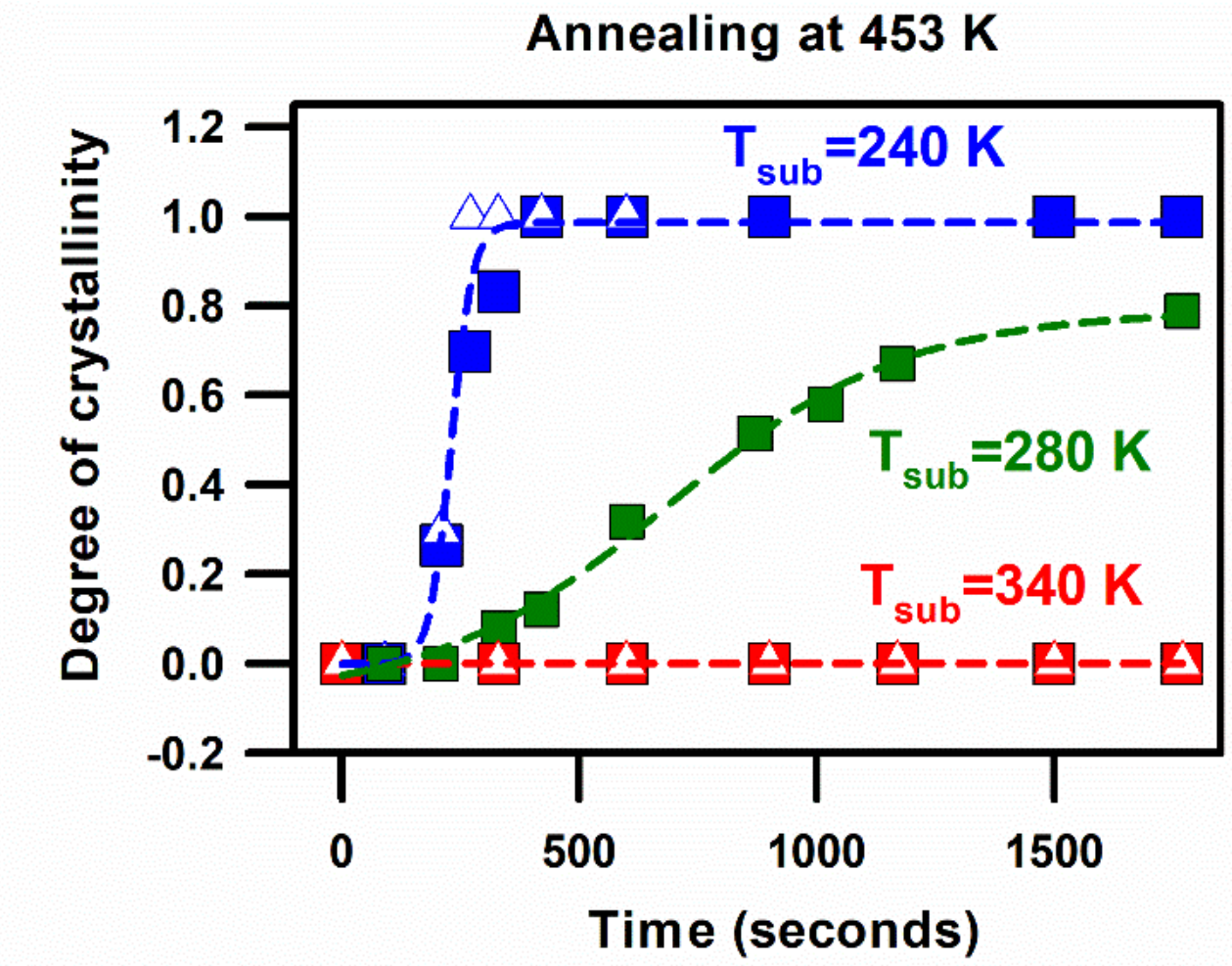

**Figure 3**: Degree of crystallinity as a function of annealing time for Alq3 glasses deposited at 240 K (blue symbols), 280 K (green symbols) and 340 K (red symbols). While the Alq3 glass deposited at 240 K (0.54 $T_g$) finishes crystallizing by ~420 s the glass deposited at 340 K (0.76 $T_g$) shows no evidence of crystallization for up to ~1800 s of annealing. The glass deposited at 280 K (0.63 $T_g$) crystallizes at a rate that is intermediate between the other two glasses. Filled and empty symbols represent measurements from duplicate samples, while the dashed lines are guides to the eye. The samples are annealed at 453 K (1.01$T_g$).

To understand the mechanism of crystallization in PVD glasses of Alq3, we study the structure of the films before the onset of crystallization. All the as-deposited glasses of Alq3 prepared for this study exhibit a broad peak along $Q_z$ at ≈ 0.8 $Å^{-1}$; this anisotropic scattering feature arises from a tendency towards molecular layering[38] and can be observed in **Figure 1** (panels a and e). When a PVD glass transforms into a supercooled liquid, the scattering becomes isotropic[17,39]. The extent of anisotropic scattering can therefore be used to monitor the transformation of a PVD glass into a supercooled liquid. To quantify the extent of anisotropic scattering at ≈ 0.8 $Å^{-1}$ we use the Hermans order parameter, $S_{GIWAXS}$, defined mathematically in equations 2 and 3 (see Methods). When $S_{GIWAXS}$=1, all the scattered intensity is localized along $Q_z$, while a $S_{GIWAXS}$= -0.5 indicates that all the intensity is concentrated along $Q_{xy}$. Isotropic packing gives rise to a $S_{GIWAXS}$=0. Shown in **Figure 4** is the $S_{GIWAXS}$ order parameter for an Alq3 glass deposited at 240 K as a function of annealing time, as well as the crystallinity. The as-deposited glass of Alq3 exhibits a $S_{GIWAXS}$ ≈ 0.07. However, after 90 s of annealing this quantity becomes approximately zero, indicating that the glass has transformed into the isotropic supercooled liquid by this time. The time required for a PVD glass to transform into a supercooled liquid is a measure of its kinetic stability and depends on the deposition temperature[17,40]. While a PVD glass of Alq3 deposited at 240 K transforms into a supercooled liquid by 90 s, as shown in **Figure 4**, the glass deposited at 340 K shows no evidence of melting into the supercooled liquid for at least ~ 2000 s (**Figure S3**).

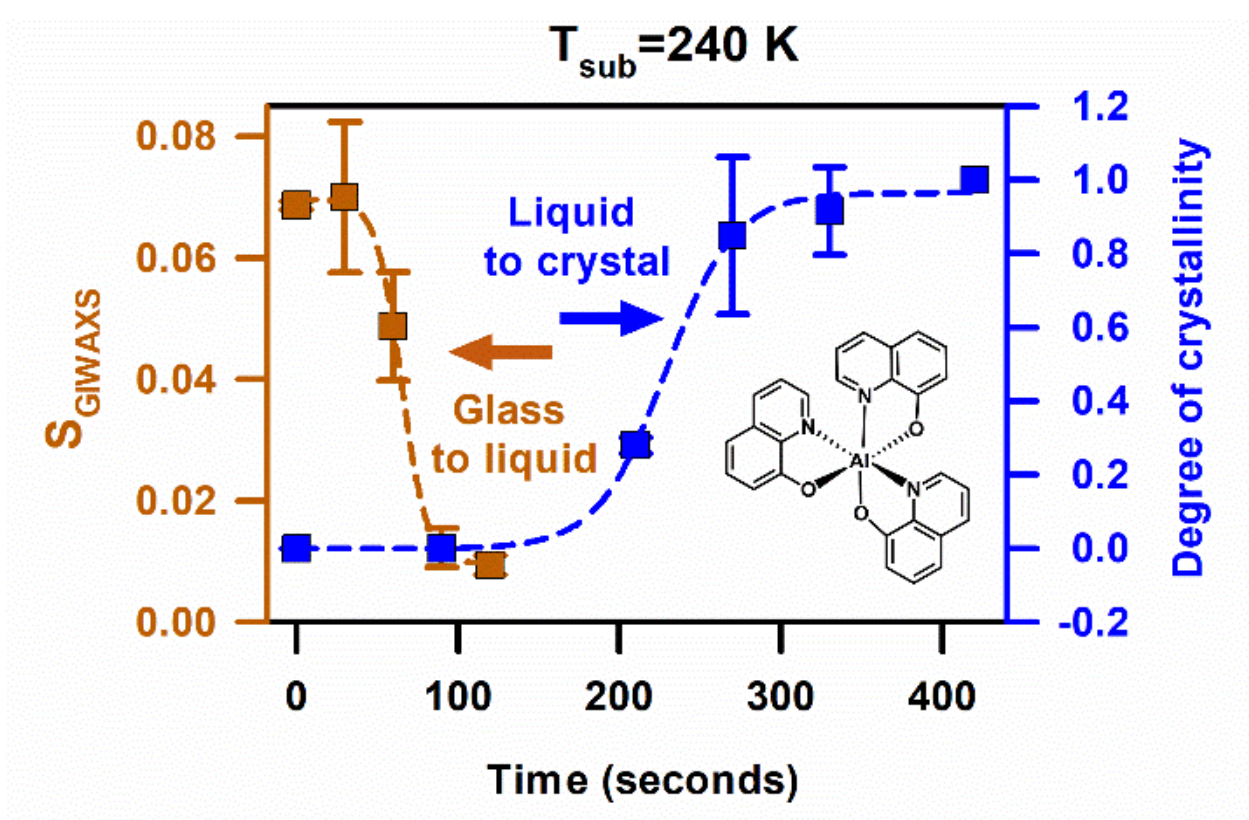

**Figure 4**: The Hermans order parameter, $S_{GIWAXS}$ (brown symbols) and the degree of crystallinity (blue symbols) as a function of annealing time for Alq3 glass deposited at 240 K. $S_{GIWAXS}$ quantifies the extent of scattering anisotropy at ≈ 0.8 $Å^{-1}$ and can be used to monitor the transformation of an anisotropic vapor-deposited glass into the isotropic supercooled liquid. The Alq3 glass deposited at 240 K transforms into a supercooled liquid by 90s and crystallizes immediately afterword. Error bars are the standard deviation from measurements on two samples and the dashed lines are eye-guides. The molecular structure of Alq3 is shown in the inset. The samples are annealed at 453 K (1.01$T_g$).

It can be seen from **Figure 4** that crystallization of a PVD glass of Alq3 follows a two-step process whereby it first transforms into a supercooled liquid and then crystallizes. The glass deposited at 240 K has completed melting into a supercooled liquid before any crystallization takes place, making it easy to simultaneously track glass melting and crystallization. One can envision a different scenario in more kinetically stable glasses where the parts of the glass that transform into a liquid crystallize, even while the rest of the glass is untransformed. Based on the data in **Figure S3** and **S4,** we expect this mechanism is applicable to the glasses deposited at 340 K and 280 K. Assessing the structural anisotropy in glasses that have partially crystallized is complicated due to peak overlap, making it difficult to simultaneously monitor glass melting and crystallization. However, from the data in **Figures 4**, **S3** and **S4**, it is clear that the glass melting kinetics is fastest in the $T_{sub}$=240 K glass and slowest in the $T_{sub}$= 340 K glass, with the $T_{sub}$=280 K being intermediate between the two. This is the same trend seen for crystallization kinetics in **Figure 3.** Thus, we conclude that the glass deposited at 340 K crystallizes the slowest because it takes the longest time to transform into a supercooled liquid.

Engineering the properties of vapor-deposited glasses to modulate crystallization is an exciting new frontier in glass science, and one that has important implications for technological applications. It has been shown for vanadia ($VO_2$)[41] and organic semiconductor CBP[11] (4,4′-bis(N-carbazolyl)-1,1′-biphenyl), that the structure of a vapor-deposited glass can influence which polymorph (or how much of a given polymorph) forms upon annealing. Our study demonstrates that the crystallization *kinetics* can also be significantly modulated by varying the glass preparation conditions. While Rodriguez-Viejo and co-workers reported a 30% slower crystallization rate in stable glasses of celecoxib compared to the liquid-cooled glass[27], we observe an order of magnitude difference in crystallization kinetics (in glasses of varying kinetic stability). As Rodriguez-Viejo and co-workers studied crystallization below the glass transition temperature, $T_g$, they observe a solid(glass)→solid(crystal) transition. In this study, crystallization is observed above $T_g$ and consequently we observe a solid(glass)→liquid→solid(crystal) transition. While the annealing temperature relative to $T_g$ is relevant for comparing celecoxib and Alq3, we expect that the more important factor is that celecoxib crystallization was limited to the glass surface; for our experiments on Alq3, bulk crystal growth must occur as the entire glass sample crystallizes by the end of the experiment.

We expect the strategy of using kinetic stability to delay crystallization to be broadly applicable to organic glass-formers including those important for technological applications such as organic semiconductors. As explained in the introduction, kinetic stability characterizes the resistance of a glass towards melting into a supercooled liquid. Kinetic stability of vapor-deposited glasses has been extensively studied, and stable glasses of more than 30 organic molecules have been identified[42], including several organic semiconductors[43,7]. Enhanced kinetic stability has already been shown to lead to higher crystallization temperatures in inorganic glass formers, such as chalcogenide[44] and metallic glasses[45].This strategy of using kinetic stability to delay crystallization will be useful for the subset of glass-formers where crystal nucleation happens at a much faster rate than glass melting, making the latter the rate determining step.

Our findings are also relevant to applications where an initial amorphous phase is used as a precursor to a desired crystalline phase. For these applications, a faster crystallization rate is desirable. Common preparation routes for thin films of crystalline phases of organic semiconductors like rubrene[46] and contorted hexabenzocoronene[47] involve post-deposition annealing of PVD glasses. Recently, Holmes and co-workers identified a strategy to form highly periodic patterns of organic semiconductors by crystallization of the as-deposited molecular glasses[48]. In solid-state epitaxy of complex oxides, an amorphous precursor phase is deposited on a single crystal template, which guides the crystallization process upon annealing[49,50]. For all these applications, it is likely that a faster crystallization rate can be achieved by depositing PVD glasses at temperatures that produce low stability glasses.

## CONCLUSION:

We demonstrate that deposition temperature significantly influences crystallization kinetics in PVD glasses of the model organic semiconductor Alq3. Our study shows that an Alq3 glass deposited at $T_{sub}$=240 K crystallizes *at least ten times* faster than one prepared at $T_{sub}$=340 K. This difference in crystallization kinetics is shown to be a result of the longer time it takes for the more kinetically stable glass to transform into the supercooled liquid, which is an intermediate in the crystallization process. We expect our findings will be useful for the broad range of organic electronics applications which require either the inhibition or the acceleration of crystallization of molecular glasses.

## SUPPORTING INFORMATION:

Figure S1 contains GIWAXS patterns for the glass deposited at 280 K. Figure S2 contains a representative time-temperature profile for the measurements shown in the main manuscript. Figure S3 and S4 show $S_{GIWAXS}$ as a function of annealing time for the glasses deposited at 340 K and 280 K, respectively. Figure S5 contains a comparison of the scattering from an annealed film of Alq3 observed in this study with the reported scattering pattern for the alpha polymorph of Alq3. Figure S6 and S7 describe how degree of crystallinity is calculated in Figure 3 and 4 of the main manuscript. Figure S8-10 contains comparison of two different methods of calculating degree of crystallinity.

## ACKNOWLEDGMENT

We acknowledge financial support from the US Department of Energy, Office of Basic Energy Sciences, Division of Materials Sciences and Engineering, Award DE-SC0002161. Data associated with this publication can be found at: https://minds.wisconsin.edu/handle/1793/75172.
We thank Yuhui Li for helpful conversations.

**TOC GRAPHICS**

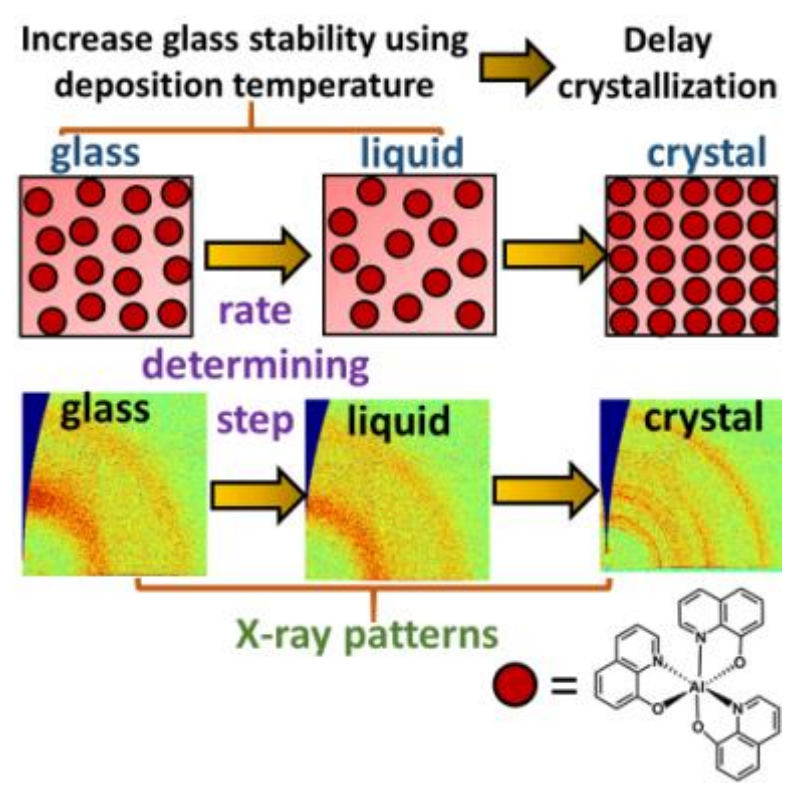

# Supplementary information

# Stable Glasses of Organic Semiconductor Resist Crystallization

*Kushal Bagchi* [a], *Marie E. Fiori* [a], *Camille Bishop* [a], *M.F. Toney*[b,c] *and M.D. Ediger*[a*]

[a] Department of Chemistry, University of Wisconsin-Madison, Madison, Wisconsin 53706, United States

[b]Stanford Synchrotron Radiation Lightsource, SLAC National Accelerator Laboratory, Menlo Park, California 94025, United States

[c]Department of Chemical and Biological Engineering, University of Colorado Boulder, Boulder, CO 80309, United States

AUTHOR INFORMATION

Corresponding Author: M.D Ediger

*Correspondence to: M.D. Ediger

Email: ediger@chem.wisc.edu

# I. <u>Raw diffraction patterns</u>

Shown in **Figure S1** are the GIWAXS patterns at three annealing times for Alq3 glasses deposited at 240 K, 280 K and 340 K. The raw GIWAXS patterns clearly show that the glass deposited at 240 K crystallizes the fastest, while the glass deposited at 340 K is the slowest to crystallize. The glass deposited at 280 K exhibits a crystallization time that is intermediate between the other two glasses. The GIWAXS patterns provide direct visual support for the conclusions of **Figure 3** of the main manuscript.

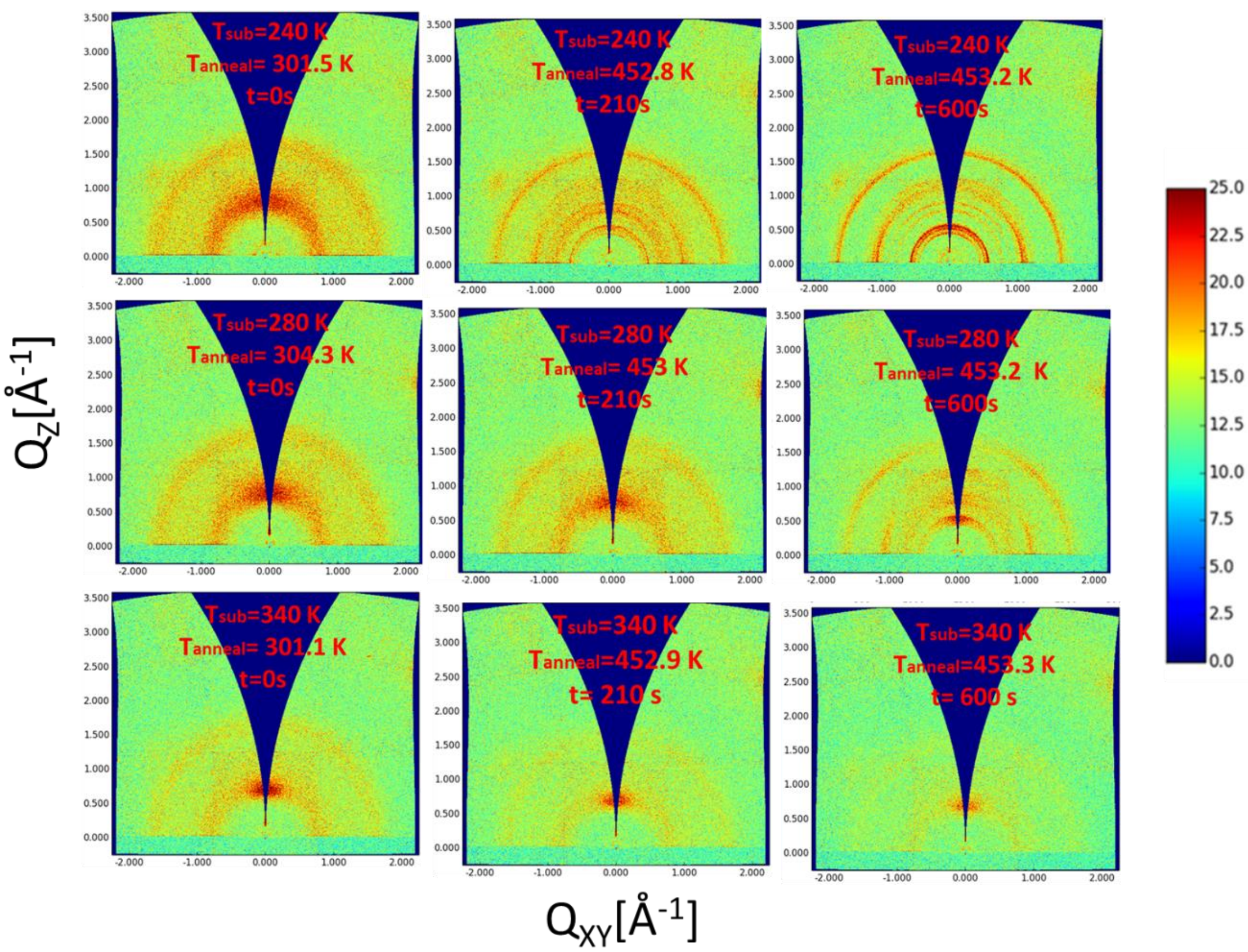

**Figure S1:** *GIWAXS patterns as a function of annealing time for Alq3 glasses deposited at 240 K(top), 280 K(middle) and 340 K (bottom). The annealing time and temperature are indicated at the top of each GIWAXS pattern. The scattering patterns for the* $T_{sub}$*=240 K and* $T_{sub}$*=340 K glasses are presented in Figure 1 of the main text and are repeated here for context.*

## II. Time-temperature profile for GIWAXS measurements

Shown in **Figure S2** is a representative time-temperature profile for the annealing measurements used in this manuscript to study crystallization kinetics. Time t=0 is defined as the start of the heating profile for this plot and for all the times reported in this paper. After 60 s of annealing, the heating stage reaches within a few degrees of the set annealing temperature.

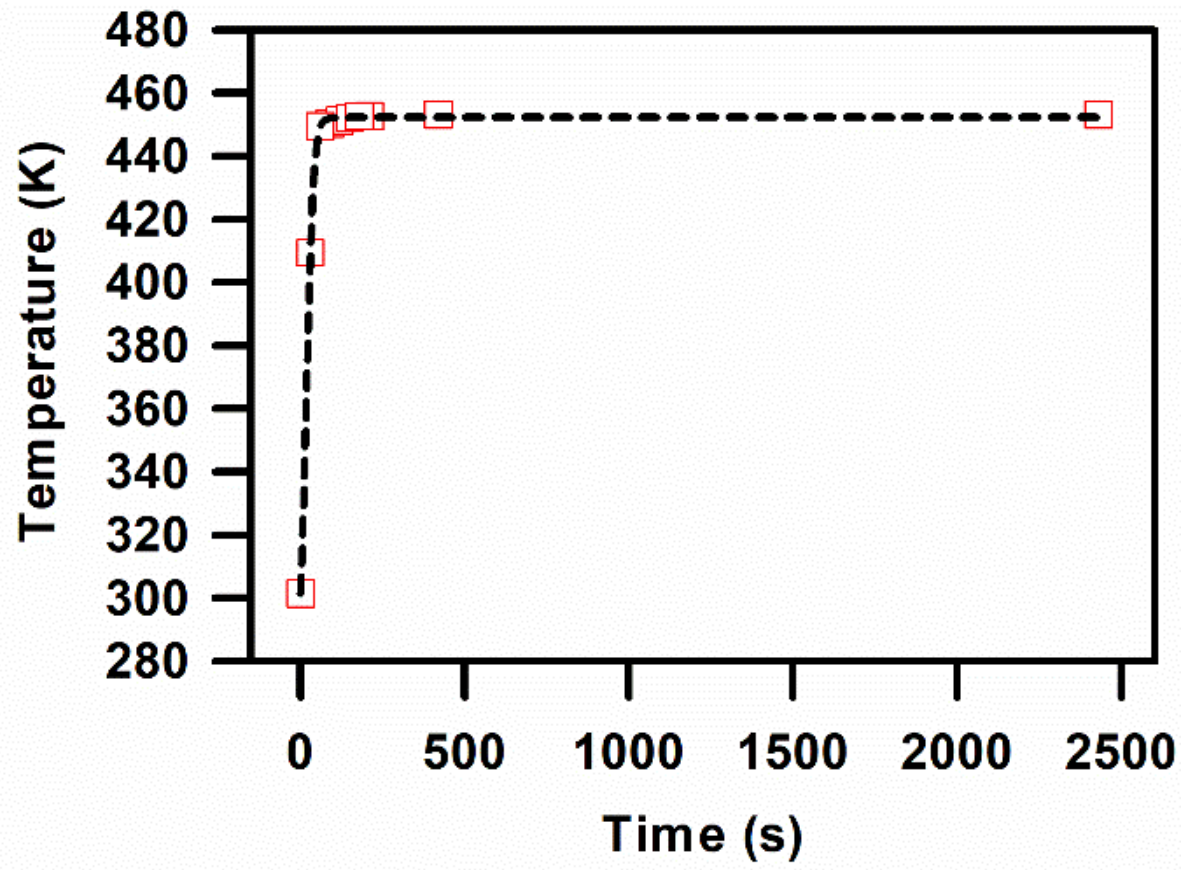

**Figure S2**: *A representative time-temperature profile for the measurements shown in Figure 1-4 of this manuscript.*

## III. Glass to liquid transformation of $T_{sub}$=340 K and 280 K glasses

Shown in **Figure S3** is the $S_{GIWAXS}$ order parameter for an Alq3 glass deposited at 340 K. The $S_{GIWAXS}$ order parameter can be used to track the PVD glass→ supercooled liquid transition; the glass deposited at 340 K shows no sign of melting into the supercooled liquid for at least 2000 s. In this time frame, the crystallinity of the sample is also close to zero (**see Figure** 3 of main manuscript). This is consistent with the mechanism discussed in the main text: the glass deposited at 340 K is more resistant to crystallization because it resists transformation into the supercooled liquid for longer times.

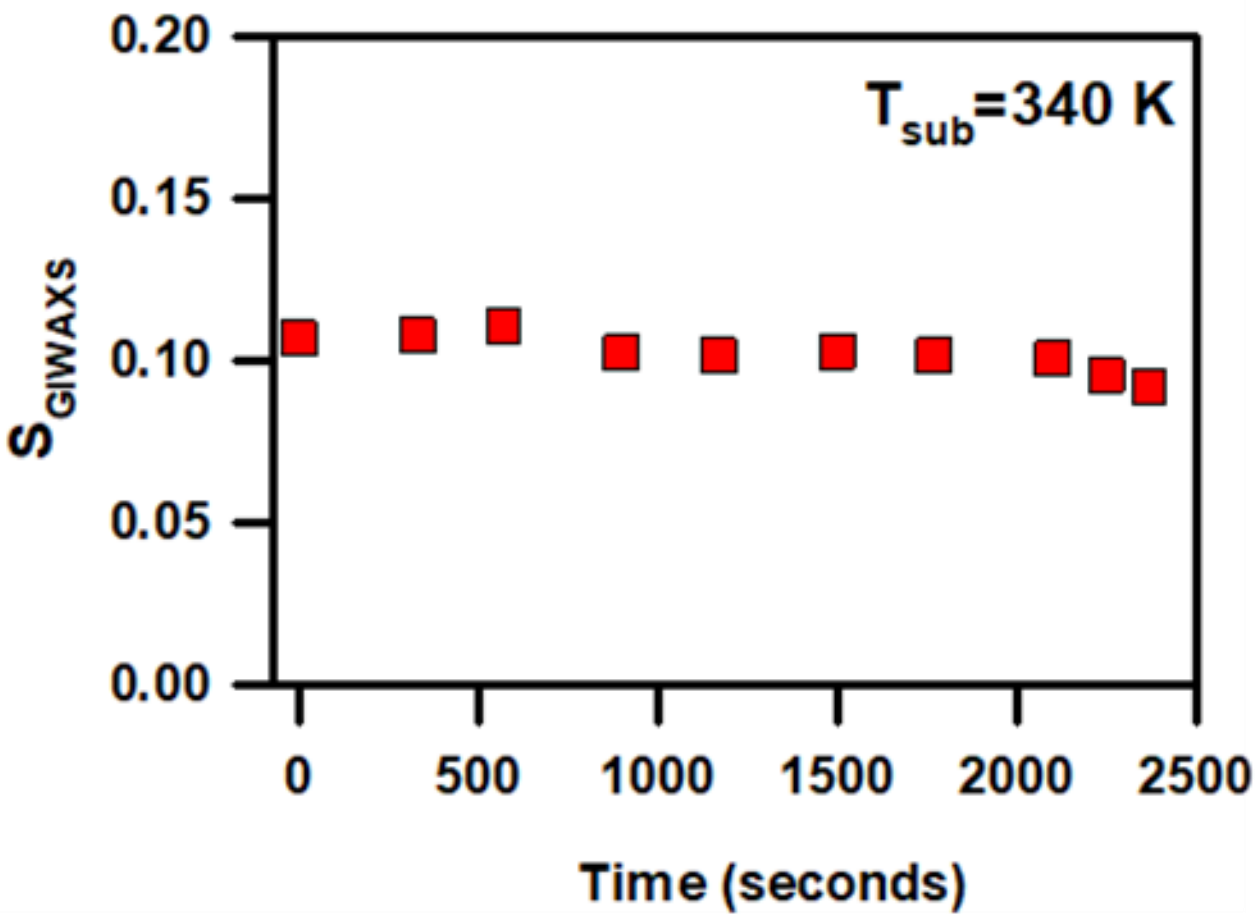

**Figure S3**: *The Hermans order parameter,* $S_{GIWAXS}$ *as a function of annealing time for Alq3 glasses deposited at 340 K. The error bars are standard deviations from measurements on two samples.*

Shown in **Figure S4** is the $S_{GIWAXS}$ order parameter for an Alq3 glass deposited at 280 K. The data is consistent with a mechanism whereby the part of the PVD glass that transforms into a supercooled liquid crystallizes, even while the remainder of the glass is untransformed. As the glass starts to crystallize however, it is difficult to measure $S_{GIWAXS}$, because the glass and crystal scatter at similar regions of reciprocal space.

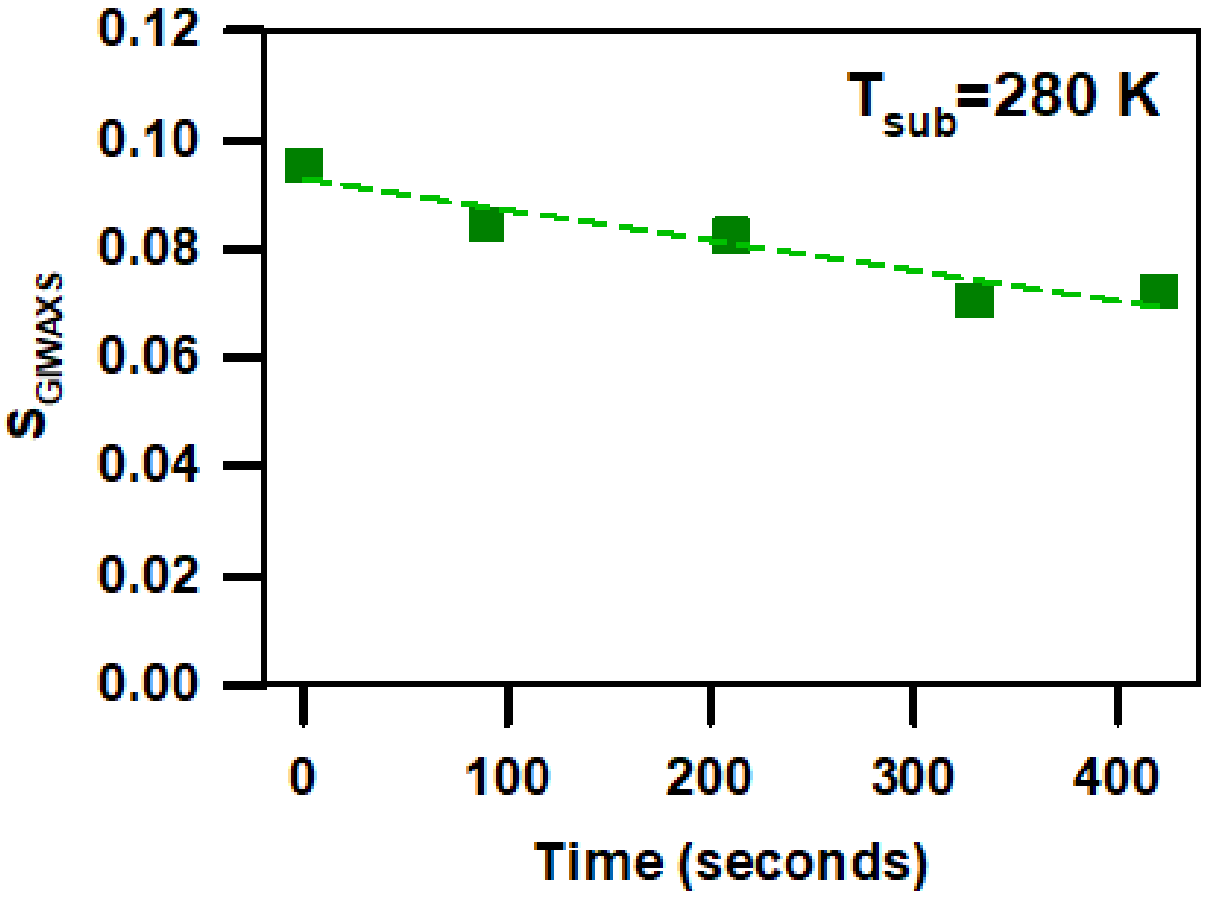

**Figure S4**: *The Hermans order parameter,* $S_{GIWAXS}$ *as a function of annealing time for an Alq3 glass deposited at 280 K.*

## IV. Polymorph identification

Shown in **Figure S5** is the scattering pattern observed in this manuscript from annealing Alq3 films compared to the literature pattern for the alpha polymorph.

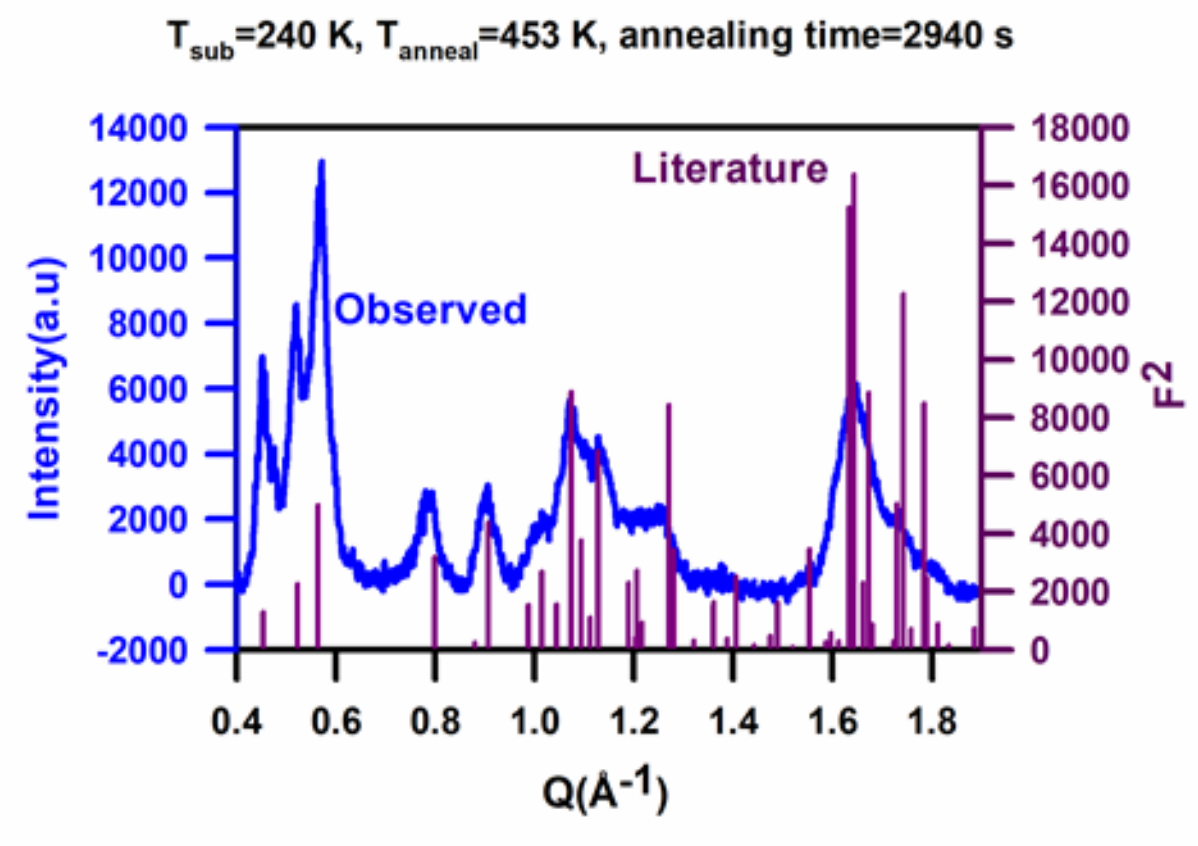

**Figure S5**: *Comparison of the scattering pattern from a crystalline Alq3 film obtained in this manuscript (blue) with the reported pattern for the alpha polymorph (dark pink). The crystallized film was initially a vapor-deposited glass prepared at 240 K; the film was annealed at 453 K for 2940 s.*

## V. Calculation of degree of crystallinity

Described below is the method used to calculate degree of crystallinity in Figure 3 and 4 of the main manuscript. For all the samples, the amorphous contribution to the total scattering is determined. The degree of crystallinity is then determined using the equation:

$$Degree\ of\ crystallinity = \frac{Total\ area - amorphous\ area}{Total\ area} = \frac{Area\ under\ crystalline\ peaks}{Total\ area}$$

A fully amorphous sample, at time t=0 s is fit to determine a mathematical description of the amorphous scatter. It is found that the amorphous scatter (for all the glasses studied) can be described as a sum of three gaussians (and an offset term), as shown in **Figure S6**.

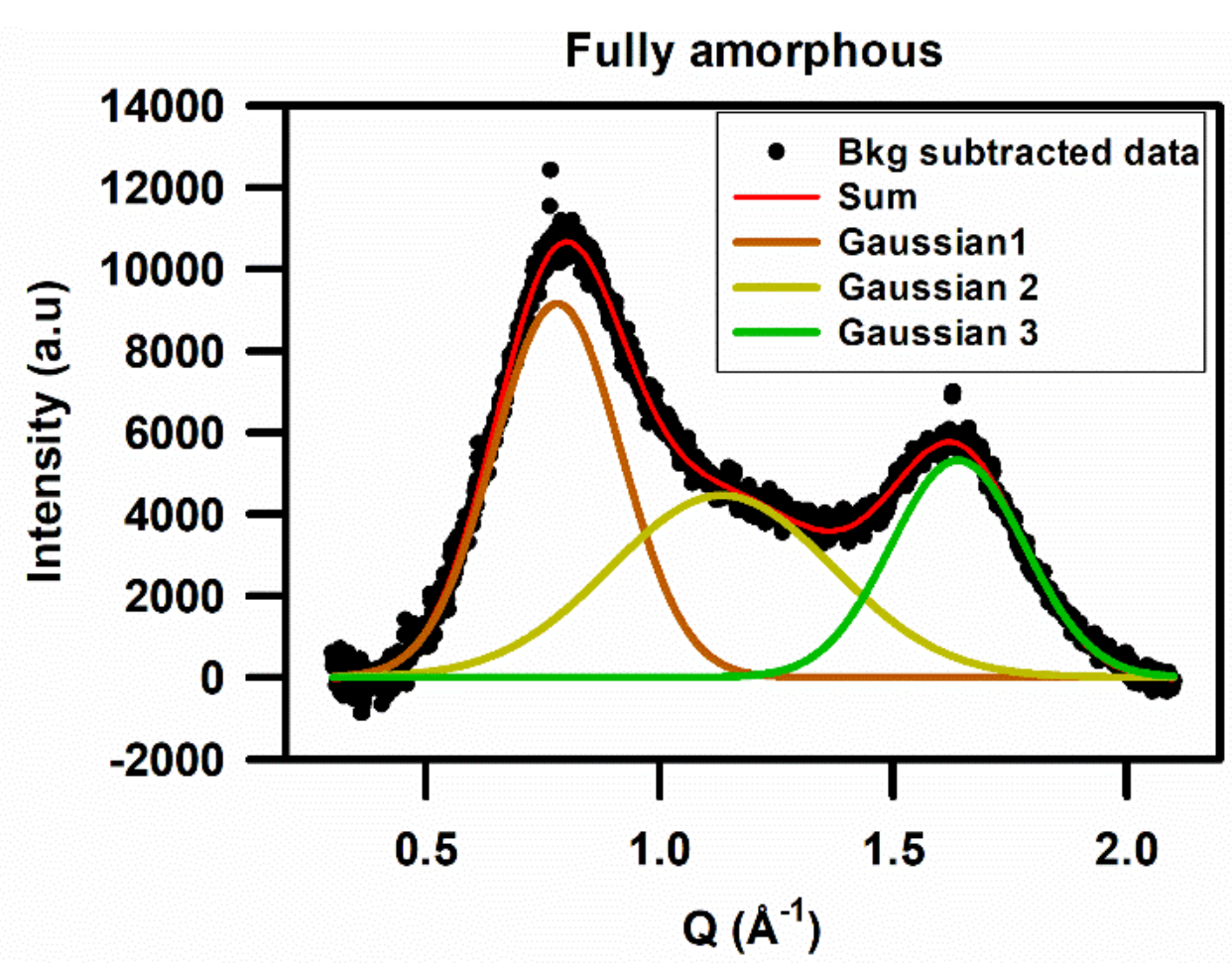

***Figure S6:*** *Fit to the scattering from a fully amorphous film. The amorphous scattering is described as a sum of three gaussian functions.*

For all subsequent fits, the peak positions, peak widths, and relative heights of the Gaussians are constrained to the values obtained from the t=0 s fit. Only the amplitude of the entire function is allowed to vary. For a semi-crystalline film, the regions where the crystal does not scatter are selectively fit, to determine the amorphous contribution as shown in **Figure S7**.

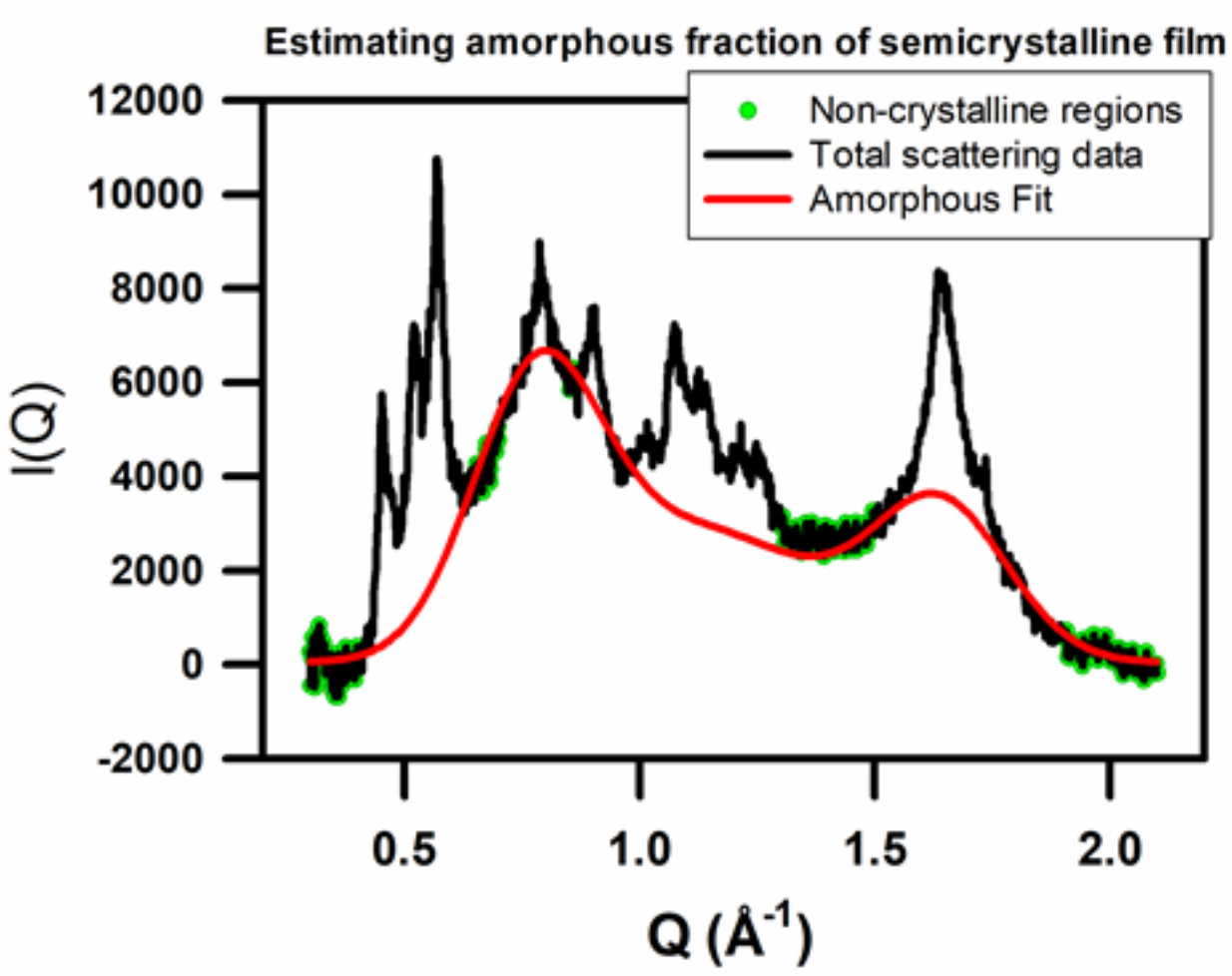

***Figure S7****: Determination of amorphous fraction for a semi-crystalline film. The regions where the crystal does not scatter(green) are fit to the function obtained from the fit shown* ***Figure S6****. The amorphous fit is shown in red, and the background subtracted data is shown in black.*

## VI. Comparison of two different methods of calculating crystallinity

To test the reliability of the procedure used in the main text for calculating degree of crystallinity we compare it below to another method. The comparisons between the two methods are shown in **Figures S9** and **S10**.

The temporal evolution of the scattered intensity between 0.4 and 0.65 $Å^{-1}$ can be used to monitor crystallization kinetics. As the initially amorphous film crystallizes, the fraction of scattered intensity between these limits increases as shown in **Figure S8 (a)**.

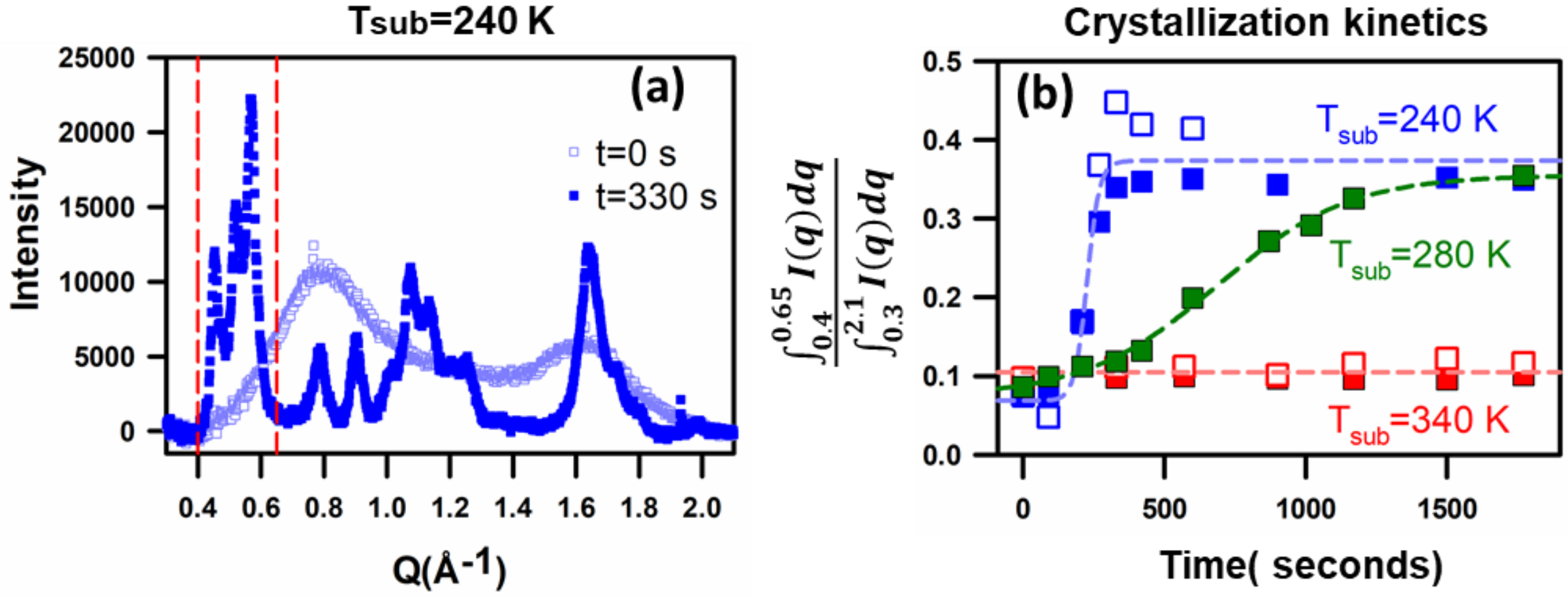

***Figure S8****: (a) X-ray scattering profiles for an Alq3 glass prepared at 240 K before annealing (t=0s) and after being annealed at 453 K for 330 s (filled symbols) and (b) crystallization kinetics for glasses deposited at 240 K (blue symbol), 280 K (green symbols) and 340 K (red symbols). Filled and empty symbols in (b) represent measurements from different samples. Figure (a) shows that as a glass crystallizes the fraction of the total scattered intensity between 0.4 and 0.65 $Å^{-1}$ increases. When the scattered intensity between these limits is normalized by the total scattered intensity from 0.3 to 2.1 $Å^{-1}$, it can be used to monitor crystallization kinetics, as shown in (b).*

The data in **Figure S8 (b)** can be used to evaluate the degree of crystallinity using the following equation:

$$\text{Degree of crystallinity}_{\text{Method 2(SI)}} = \frac{\left[\frac{\int_{0.4}^{0.65} I(q)dq}{\int_{0.3}^{2.1} I(q)dq}(t) - \frac{\int_{0.4}^{0.65} I(q)dq}{\int_{0.3}^{2.1} I(q)dq}(t=0s)\right]}{\left[\frac{\int_{0.4}^{0.65} I(q)dq}{\int_{0.3}^{2.1} I(q)dq}(t=\infty) - \frac{\int_{0.4}^{0.65} I(q)dq}{\int_{0.3}^{2.1} I(q)dq}(t=0s)\right]}$$

The denominator in the integrals normalizes for temporal fluctuations in scattered intensity. To approximate the term $\frac{\int_{0.4}^{0.65} I(q)dq}{\int_{0.3}^{2.1} I(q)dq}(t=\infty)$, the asymptotic value from a sigmoidal fit is applied.

To distinguish the method described, from the one used in the main text it shall be called "Method 2 (SI) ". The method used in the main text is labelled "Method 1(Main)" . Comparison of the two methods are shown below in **Figures S9** and **S10**.

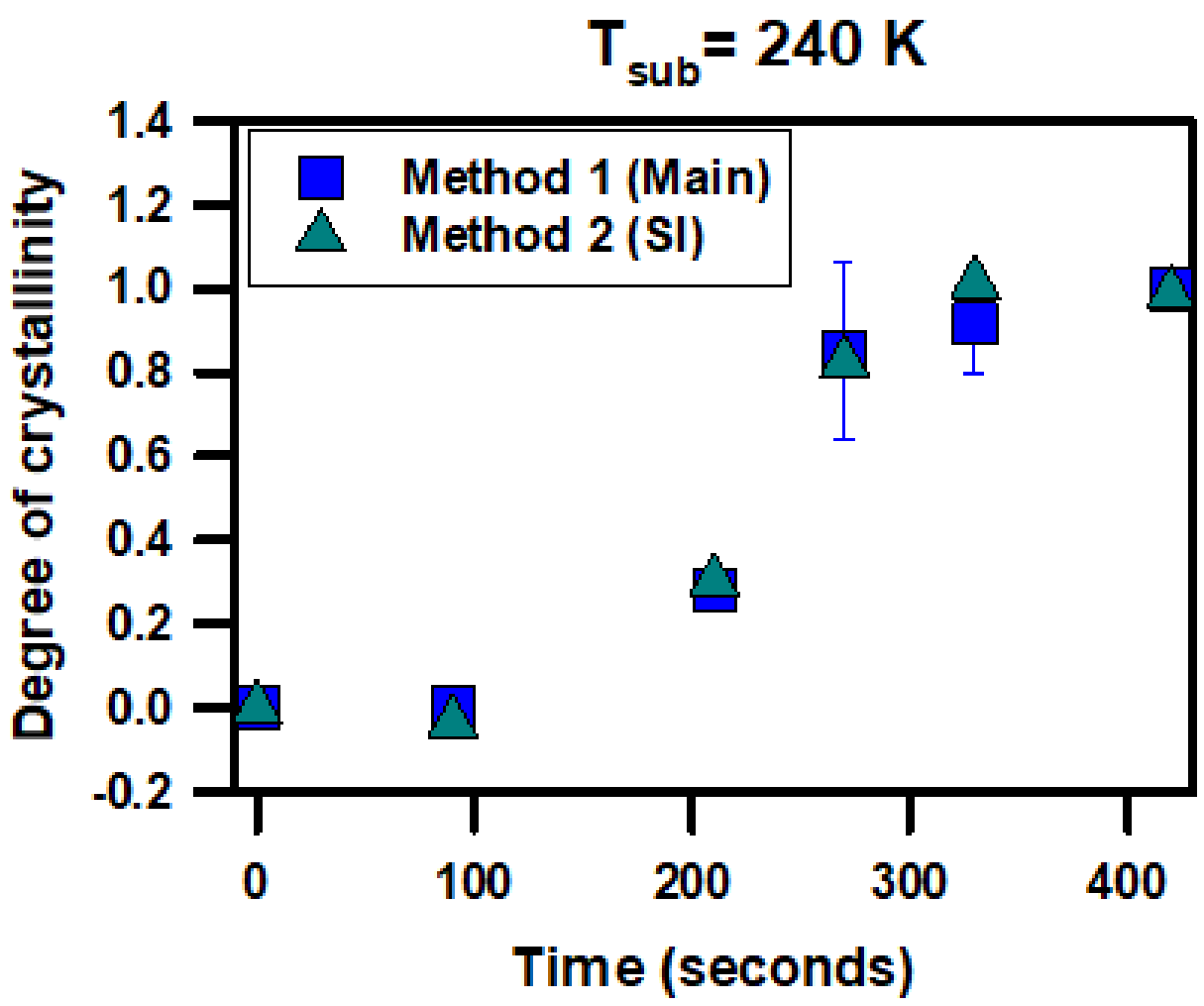

***Figure S9****: Comparison of two different methods of calculating crystallinity for the $T_{sub}$=240 K samples. The error bar is the standard deviation from two duplicate samples. Method 1(Main) is used in the main text and is described in section V of the SI, while Method 2 is described in section VI of the SI. Agreement between the two methods is good.*

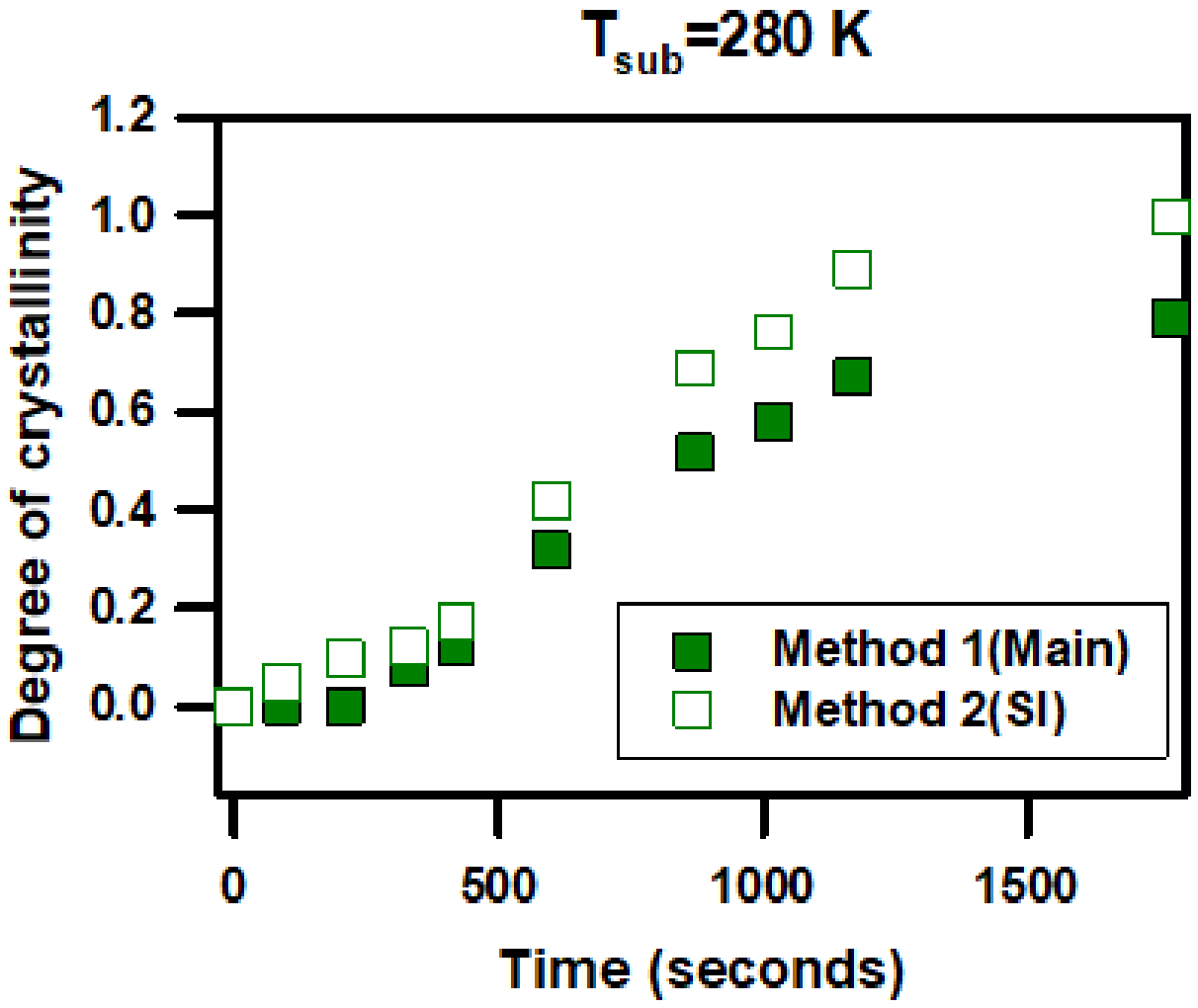

***Figure S10****: Comparison of two different methods of calculating crystallinity for the $T_{sub}$=280 K glass. Method 1(Main) is used in the main text and is described in section V, while Method 2 is described in section VI of the SI. Agreement between the two methods is good.*